\documentclass[12pt]{revtex4}

\usepackage{amsmath,graphicx,latexsym}
\usepackage{bm}
\usepackage{epsfig}
\usepackage{amsmath}
\usepackage[colorlinks,linkcolor=blue,citecolor=red]{hyperref}
\usepackage{epsfig}
\usepackage{latexsym,amsmath,amssymb}
\usepackage{graphicx}
\usepackage{slashed}
\usepackage{bm}
\usepackage{epsfig}
\usepackage{dcolumn}

\usepackage{array}
\usepackage{booktabs}
\usepackage{tabu}
\def\beq{\begin{equation}}
\def\eeq{\end{equation}}
\def\ber{\begin{eqnarray}}
\def\eer{\end{eqnarray}}
\def\benu{\begin{enumerate}}
\def\eenu{\end{enumerate}}

\def \lleq {\lower0.9ex\hbox{ $\buildrel < \over \sim$} ~}
\def \ggeq {\lower0.9ex\hbox{ $\buildrel > \over \sim$} ~}

\begin{document}

\title{Constraints on warm power-law inflation in light of Planck results}

\author{Zahra Ghadiri}
\email{zghadiry@gmail.com}
\affiliation{Sanandaj Branch, Islamic Azad university, Sanandaj, Iran}
\author{Ali Aghamohammadi}
\email{a.aqamohamadi@gmail.com;a.aghamohamadi@iausdj.ac.ir}
\affiliation{Sanandaj Branch, Islamic Azad university, Sanandaj, Iran}
\author{Abdollah Refaei}
\email{abr412@gmail.com;A.Refaei@iausdj.ac.ir}
\affiliation{Sanandaj Branch, Islamic Azad university, Sanandaj, Iran}
\author{Haidar Sheikhahmadi}
\email{h.sh.ahmadi@gmail.com;h.sheikhahmadi@ipm.ir}
\affiliation{School of Astronomy, Institute for Research in
Fundamental Sciences (IPM), P. O. Box 19395-5531, Tehran, Iran}
\affiliation{Center for Space Research, North-West University, Mafikeng, South Africa}

\begin{abstract}
The constraints on a general form of  the power-law potential and  the dissipation coefficient in the framework of warm single field inflation imposed by Planck data will be investigated. {By Considering a  quasi-static Universe, besides a slow-roll condition, the suitable regions in which a pair of theoretical free parameters are in good agreement with Planck results will be estimated}. In this method instead of a set of free parameters we can visualize a region of free parameters that can satisfy the precision limits on theoretical results. On the other side, when we consider the preformed quantity for the amplitude of scalar perturbations, the conflict between obtained results for free parameters in different steps dramatically will be decreased. {As have done in prominent} literature, based on the friction of the environment,  we can divide the primordial Universe to the two different epochs namely weak and strong dissipative regimes. For the aforementioned eras, the free parameters of the model will be constrained and the best regions will be obtained. To do so, the main
inflationary observables such as tensor-to-scalar ratio, power-spectra of density perturbations and gravitational waves, scalar and tensor spectral indices, running
spectral index and the number of e-folds in both weak and strong regimes will be obtained. Ultimately, it can be visualized, this model  can make concord between theoretical results and  data originated from cosmic microwave background and Planck $2013$, $2015$ and $2018$.
\\
\textbf{PACS numbers:} 98.80.Cq; 04.20.CV\\
\textbf{keywords:}Warm power-law inflation, Weak and strong dissipative regimes, Planck $2013$ and $2015$
\end{abstract}

\maketitle


\section{Introduction}\label{secintro}
In this work by virtue of the Planck results \cite{Ade:2013ktc,Ade:2013uln,Ade:2013ydc,Planck2015,Planck2018} the free parameters of  both the  power-law potential and   the dissipation coefficient in the framework of warm inflation will be constrained.\\

After four decades of introducing, {now the inflation  can}  be considered as the most  acceptable paradigm to explain  the evolution of early Universe \cite{Sta80,Gut81,Lin82,Lin83,Lin86a,Lin86b,Alb}. The main {achievement} of this  theory goes back to cope with three well-known problems of the standard big bang theory namely the horizon, flatness and  monopoles problems \cite{Lid00,Bas,Lem,Kin,Bau09,Bau14}. Even better,  {the} inflation theory  provides a mechanism to explain structure formation and the source of the observed anisotropies in the Cosmic Microwave Background (CMB) radiation \cite{Larson,Bennettc,Jarosik}. Besides, this scenario {could assess}  a correct result for the amplitude of primordial perturbations compared to observations
\cite{Bas,Liddle0,Langl,Lyth,Lyth1,Guth00,Lidsey97,Mukhanov-etal,Haidar0,Haidar,Haidar2}.
{The} various inflationary models have been investigated in \cite{Robert,Lid00}, that could be divided into {two well-known approaches}  as super-cold and warm inflationary proposals. In the warm inflation scenario \cite{Taylor,Oliveira,Oliveira2,6252603,62526,Bastero,Cid,Luca},{ during the initial rapid expansion of the Universe, the continuous radiation production based on thermal perturbations is arisen and so the interaction between  radiation and other components of the extra hot primordial soup can be visualized} \cite{Oliveira,Oliveira2}. One important result from such interactions is that the energy density  evolutions {could} be considered almost constant \cite{Taylor}. Thanks to such evolutions the inflationary phase smoothly enteres into radiation epoch, without{ introducing}  pre-heating and reheating phases which are critical parts of super-cold inflation models.  In other word,  {The remedy of the warm inflation } to get rid of the graceful exit problem  which super-cold inflation was faced,  could be considered as an advantage, for more details {we refer the reader to } \cite{Oliveira2,Antonella,warm,warm0,warm1,warm2,28,Karami1,Abol01}. One way to show the interaction between different components of the primordial thermal bath, i.e. inflationary era,  is invoking  the dissipation effect as a rule of the flow of energy between them \cite{Fang}. Consequently, such interactions cause  a high friction environment and  the friction term,{ i.e. $\Gamma {\dot{\phi}}^2$, that} will be appeared in the conservation equations of scalar field and radiation, for a precision study one can see \cite{Moss}. As a supplementary discussion, a principal condition in order to happen the warm inflation  is {that} the radiation temperature  satisfy condition $T>H$, where $T$ and $H$ are temperature and Hubble parameter respectively \cite{Herrera:2015aja,Herrera:2015a1,Herrera:2015a2}. In  Refs. \cite{warm,warm0,warm1,warm2}, it was noticed that the  quantum fluctuation and thermal perturbations{,} could be dependent upon the parameters $T$ and $H$ respectively. According to the condition $T>H$, the thermal fluctuations play a crucial role in producing  the primary density perturbations, as a seeds for large-scale structure formation. In this case, {the thermal fluctuations  entirely dominated in  comparison to the quantum portion in primordial fluctuations} \cite{warm0,warm1,warm2,Berera2}.\\
On the other hand, in 1983, A. Linde has introduced a new proposal for inflation based on the chaotic scalar field namely chaotic inflation \cite{Lin83}. In this theory he put { forward} the power law potential , as same as  {the }quantum field theory, and it plays the role of  corner stone of his theory. Hence the power law {potentials received a lot of interests}, because of their simplicity and compatibility with observations and also solving the graceful exit problem of old and new inflation. But the answer of this simple query that{,} which exponents are able to run warm inflation in a better way compared to Planck results is remained  ambiguous  \cite{Lyth}. Therefore our main motivation to start this study {is} finding the answers of the aforementioned question. Technically, for both regimes we will calculate the regions which give the best estimated values for free parameters of our model.  After that, we will examine the accuracy of our estimations for some pares of best fitted parameters compared to our criterion, {i.e. } Planck $2013$, $2015$ and {$2018$ } constraints on inflationary data \cite{Ade:2013ktc,Ade:2013uln,Ade:2013ydc,Planck2015,Planck2018}.\\
This paper is organized as follows:
 In Section~2, the main dynamical equations of warm inflation will be expressed and so the inflationary parameters will be evaluated. Additionally, Section~3 will be devoted to study on the power-law potentials in a weak dissipative regime and the region which contains the best fitted free parameters of the model will be appeared.  And Section~4 will be dedicated to investigate the behaviour of power-law  potentials in the strong dissipative regime. Similar to  the weak regime,  the parameters which can derive warm inflation will be estimated by virtue of  the Planck results. Ultimately, Section~5 is devoted to conclusion and final remarks.

\section{General framework}\label{warmrew}
There is a spatially flat Friedmann-Lima\^{i}tre-Robertson-Walker (FLRW) space time with signature $-2$. We assume that the Universe consists of two different {\bf components},  in which one of them can be a self-interacting scalar field  $\phi$ and {all the remnant } components  are considered as perfect fluid as well. The energy density and pressure of such scalar field are expressed  as follows:, respectively,
$$\rho_{\phi}=\frac{\dot{\phi}^2}{2}+V(\phi)~,$$
 and
$$P_{\phi}=\frac{\dot{\phi}^2}{2}-V(\phi)~.$$
The energy density of the aforementioned perfect fluid, i.e. almost radiation, is presented as $\rho_r$.
Also the first Friedmann equation could be expressed as follows:
\begin{equation}
H^2=\frac{1}{3M_{p}^{2}}(\rho_{\phi}+\rho_r)~,\label{Freq}
\end{equation}
where $M_{p}^{2}=\frac{1}{8\pi G}$ is the well-known reduced Planck mass and is {of the} order of $10^{18} GeV$.
The evolution equation of scalar field in warm inflation{ can be } expressed as \cite{warm0,warm1,warm2,Berera2,sayar}
\begin{equation}\label{ddotphi}
\ddot \phi  + 3H(1 + Q)\dot \phi  + V' = 0~,
\end{equation}
where the parameter $Q\equiv \Gamma/3H$ is introduced as an anomalous dissipation function,{ that} should be determined,  and $\Gamma$ is
 the  dissipation coefficient, usually can be {introduced} as an ansatz.
The conservation equations of $\rho_{\phi}$ and $\rho_{r}$ are described by the following equations \cite{warm,Berera2}
\begin{equation}\label{rhodotphi}
\dot{\rho_{\phi}}+3 H(\rho_{\phi}+P_{\phi})=-\Gamma \dot{\phi}^{2}~,
\end{equation}
and
\begin{equation}\label{rhodotr}
\dot{\rho}_{r}+4H\rho_{r}=\Gamma \dot{\phi}^{2}~, %
\end{equation}
where the  dissipation coefficient {is a positive parameter}, $\Gamma>0$. From above equations it is understood that the flow of energy is from scalar field to the radiation. We can consider an ansatz for  dissipation coefficient $\Gamma(T,\phi)$ as
\begin{equation}
\Gamma(T,\phi)=aT^n\phi^{1-n}, \label{Gamma}
\end{equation}
here {the  parameter $T$} is the  temperature of the fluid and for the certain reasons the dissipation coefficient can be introduced as a function of temperature and scalar field
\cite{Zhang:2009ge,BasteroGil:2011xd,BasteroGil:2012cm,BasteroGil:2012cm1,warm,Visinelli}.
{In fact, the main motivation is taken from \cite{sdel, sdel2,26,BasteroGil:2011xd,BasteroGil:2012cm,BasteroGil:2012cm1,Herrera:2015a2}, in which $n=1$ correspond to $\Gamma\propto T$, that is describing the high-temperature super symmetry case; for $n=0$, the $\Gamma$ is only depends on the scalar field, $\Gamma\propto\phi$ which imparts an exponentially decaying propagator in the super symmetry case; and for $n=-1,$ there is $\Gamma\propto T^{-1}\phi^{2}$, which is consistent with the non-supersymmetry case.  As well, the case $n=3$, $\Gamma=T^3\phi^{-2}$   was investigated in \cite{sdel2}.   }
During warm inflation, the energy density of scalar field  is the dominant component in comparison to the radiation one i.e. $\rho_\phi\gg\rho_r$ {\cite{warm,warm0,warm1,warm2}}. In other words, the expansion rate  is smaller than  the radiation energy density, i.e.    $\rho_r^{1/4}>H$ or $T>H$,  which is a critical requirement of happening a healthy warm inflation.
Besides, if we consider the  notion of the slow roll conditions, it is assumed that during inflation the radiation production becomes quasi-stable i.e. $\dot{\rho
}_r\ll4H\rho_r,\Gamma\dot{\phi}^{2}$
 {\cite{warm,warm0,warm2}} and according to (\ref{ddotphi}) one concludes $\ddot \phi \ll3H(1 + Q)\dot \phi$. Thence, the equations (\ref{ddotphi}) and  (\ref{rhodotr}) can be approximated as
\begin{eqnarray}
\rho_r=\alpha T^4\simeq \frac{\Gamma\,\dot{\phi}^{2}}{4H}~,\label{rhor}\\
3\,H\,(1+Q)\dot{\phi}\simeq -V_{,\phi}~.
\label{phidot}%
\end{eqnarray}
where $\alpha=\pi^2 g_\star / 30$ is the {Stefan-Boltzmann} constant, {that, the number of degrees of freedom of the radiation field  is $g_\star=228.75$ in the Minimal Supersymmetric Standard Model (MSSM), }   {hence for the parameter $\alpha$ we receive $\alpha=70$} .
The necessary condition for inflation in the warm inflation context, is that the slow-roll parameters Eq.\eqref{srparam} obey Eq.\eqref{srcon}
\begin{equation}
\epsilon=\frac{M^2_p}{2}\left(\frac{V_{,\phi}}{V}\right)^2,\,\,\,\eta=M^2_p\left(\frac{V_{,\phi \phi}}{V}\right)~,\label{srparam}
\end{equation}
where the slow-roll conditions for warm inflation must  satisfy the following conditions \cite{warm2,26,Grigorios,R2,R3,R4,R5}
\begin{equation}
\epsilon \ll 1+Q,\,\,\,\eta \ll 1+Q.\label{srcon}
\end{equation}
Based on the definitions of slow-roll parameters  whenever one of these parameters becomes equal to  $1+Q$ the inflation process is terminated. Another important parameter to describe the inflationary
evolution is the number of e-folds. This parameter plays an important role in solving the horizon problem and  is defined as
\begin{equation}
N =  - \int_{{\phi _{end}}}^\phi  {\mkern 1mu}  \frac{H}{{\dot \phi }}{\mkern 1mu} d\phi.~\label{Nefold}
\end{equation}
Additionally, to estimate ,best values for, the free parameters of the model we have to obtain the amplitude of scalar and tensor perturbations, tensor-to-scalar spectrum ratio, the scalar spectral index and running parameter which are expressed respectively as follows.
{Before going forward, we should emphasise  we considered a much more accepted expression for the amplitude of the primordial spectrum \cite{MB, MMV}.}
\begin{equation}
{{\cal P}_s}  = \frac{{25}}{4}\frac{{{H^2}}}{{{{\dot \phi }^2}}}\delta {\phi ^2}~,\label{ps}
\end{equation}
\begin{equation}
{{\cal P}_t}  =  \frac{{2{H^2}}}{{{\pi ^2}{M_p}^2}}~,\label{pt}
\end{equation}
\begin{equation}
\label{rratio}
r \equiv \frac{{{{\cal P}_t}}}{{{{\cal P}_s}}}~,
\end{equation}
\begin{equation}
\label{ns}
{n_s} - 1 \equiv \frac{{d\ln {{\cal P}_s}}}{{d\ln k}}~,
\end{equation}
and
\begin{eqnarray}
\label{dnsk}
\alpha_s=\frac{{d{n_s}}}{{d\ln k}}~.
\end{eqnarray}
From the above equations and slow roll conditions  it {is understood}  that at the sound horizon exit , i.e. $aH = c_s k$, one can obtain \cite{Unn13}
\begin{equation}
\label{exithorizon}
\frac{{\rm{d}}}{{{\rm{d ln k}}}} \simeq -\frac{{\rm{d}}}{{{\rm{dN}}}}~,
\end{equation}
in above relation, cause of slow roll notion, the parameters $H$  and  sound speed $c_s$ are considered as constant.\\
In the following, the calculations will be divided  into two different sections namely the weak and strong
dissipation regimes, i.e. $Q\ll 1$ and $Q\gg 1$ respectively. Because of the importance of power-law potentials in deriving inflation  we are going to consider a general form of the power-law potentials and investigate the behaviour of such  potentials in the framework of warm inflation scenario. {In short, for  the weak dissipation regime we shall see that  the parameter  which drives  inflation is the Hubble parameter.} But in the strong regime, the situation {because} of {the high friction environment}, is completely different and the evolution of inflation is operated by the dissipation coefficient $\Gamma$. The impact of different types of dissipation coefficients were investigated extensively in the literature  and we refer the reader to \cite{BasteroGil:2012cm,BasteroGil:2012cm1,Herrera:2015aja}. For both regimes we  interested in finding the best fitted free parameters of the model especially for  the exponents of variables of   potential and dissipation coefficient functions.

\section{Consistency Of Power-Law Potential In Weak dissipative Regime }\label{weakpl}
\begin{figure}[ht]
\centering
\includegraphics[scale=.43]{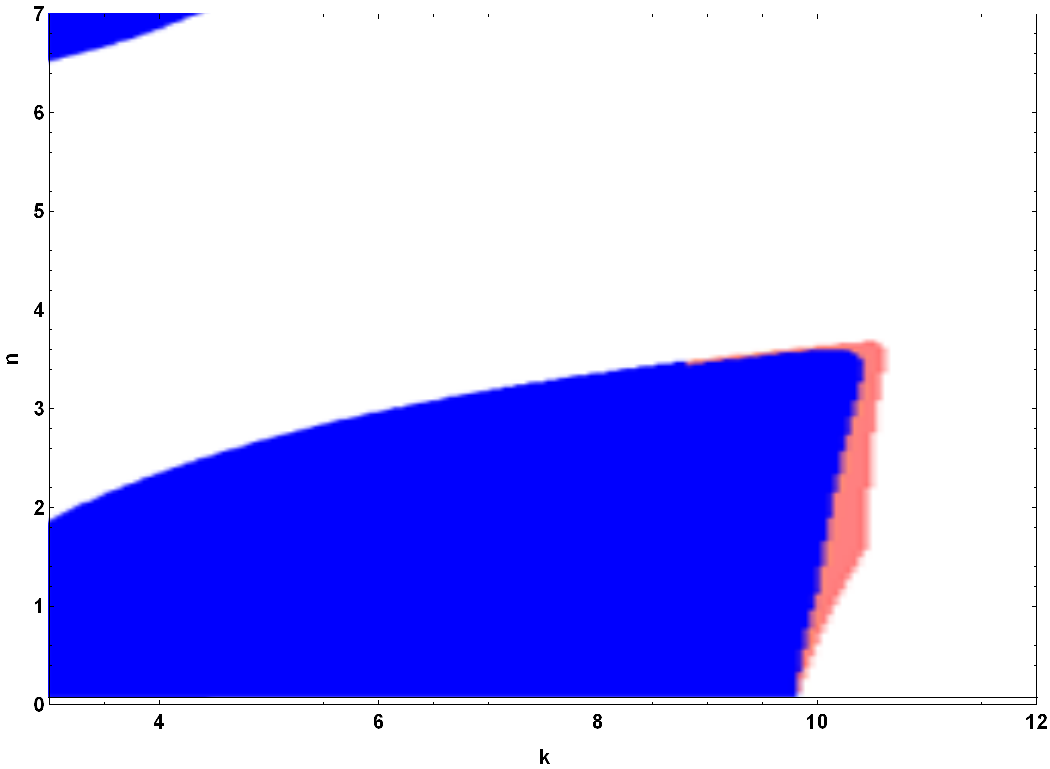}
\caption{{\it{This figure shows a set of {pair} $(n, k)$  that can give the best fitted values based on the  $(r-n_s)$ diagram originated from Planck $2015$ observational {data}. Here the dark blue color shows the results for $68\%$ CL of Planck $2015$ data, and the light red color shows those, that put the result in $95\%$ CL. To plot this diagram we used $V_0=2.7\times10^{-25}$, and $M_p=1$..}}}
\label{fignrweak}
\end{figure}

In this section, given that in this section our model evaluates based on the weak dissipation regime in  which $Q\ll1$,  by introducing the potential as
 \begin{equation}\label{Potential}
\ V(\phi)={V_0}\,{\phi^{k}}~,
 \end{equation}
{and imposition }slow-roll approximations, the Eqs. (\ref{Freq}) and (\ref{ddotphi}) take the following form
\begin{equation}
\ H^2=\frac{1}{3M^2_p}{\ V(\phi)} \label{WeakFreq}~,%
\end{equation}
and
\begin{equation}\label{phidotweak}
 3H\dot \phi  + V_{,\phi} = 0~,
\end{equation}
where  subscript   $(,\phi)$ stands for derivative with respect to $\phi$.
In Eq.(\ref{rhor}) the temperature of the
radiation according to the Eqs. (\ref{Gamma}) and (\ref{Potential}) obtained as
\begin{equation}\label{TeWeak}
T=\gamma {\phi ^{\frac{{2 - k + 2n}}{{2( - 4 + n)}}}}~,
\end{equation}
where
 $\gamma={(\frac{{a{k^2}{M_p}^3\sqrt {{V_0}} }}{36 \alpha })^{\frac{1}{{4 - n}}}}~.$
In addition, the first slow-roll parameter (\ref{srparam}) for this regime reduces to
\begin{equation}
\epsilon=\frac{{{k^2}{M_p}^2}}{{2{\phi ^2}}}~.\label{srw1}
\end{equation}
Generally, the inflation period terminates when $\epsilon=1$, where using Eq.(\ref{srw1}), one achieves $\phi_{\textup{end}}=\frac{{k{M_p}}}{{\sqrt 2 }}$.
Besides, from  Eq.(\ref{Nefold}) the value of scalar field at the time  of the exit of the  horizon is given by the following equation
\begin{equation}
\phi  = \sqrt {2k{M_p}^2N + {\phi _{end}}^2}~.\label{phiexit}
\end{equation}
In the  following, we want to calculate the perturbation parameters based on our model and  {compare} them with observational data. To do so, we consider  Eq.(\ref{ps}) in  which for weak regime  leads  $\delta\phi^2\simeq H T$. Whereas the  fluctuations in the warm inflationary theory are created
by the thermal fluctuations instead of the quantum fluctuations, \cite{Herrera:2015aja}, accordingly the amplitude of scalar perturbation becomes
\begin{equation}
{\cal P}_s = \frac{{25\gamma }}{{4{k^2}{M_p}^5}}{(\frac{{{k^2}{M_p}^2}}{2} + 2k{M_p}^2N)^{\frac{{14 + 5k - nk -6n  }}{{4(4 - n)}}}}\sqrt {\frac{{{{\rm{V}}_0}}}{3}}~.\label{psN}
\end{equation}
From Eq.(\ref{psN}) and definition of $\gamma$ we can express $a(n, k, N)$ as follows:
\begin{eqnarray}
a(n, k, N) &=& ({\sqrt{\frac{{{3}}}{V_0}}}\frac{{4{k^2}{M_p}^5}}{{25\gamma^\ast }})^{{4-n}}\\ \nonumber
&\times&{(\frac{{{k^2}{M_p}^2+ 4k{M_p}^2N}}{2} )^{\frac{{( 14 + 5k - nk -6n) }}{{-4}}}}{\cal P}_s^{4-n}~.\label{a-psN}
\end{eqnarray}
Here $\gamma^\ast={(\frac{{{k^2}{M_p}^3\sqrt {{V_0}} }}{36 \alpha })^{\frac{1}{{4 - n}}}}$ is introduced for more convenience in writing the equations and we shall consider the observational value for the amplitude of scalar perturbation as $2.17\times 10^{-9}$. Sometimes although the free parameters satisfy the observational constraints on  quantity the scalar spectral index, it could not concluded that they can make a concord between observations and theoretical results for ${\cal P}_s$. So it is better we examine the accuracy of the free parameters of the model based on Planck results as our criterion. For instance here we can introduce our free parameters of the model as $n,~k,~ N$ and other parameters can be estimated based on the behaviour of these parameters as well.
To do so, at first, from Eqs.(\ref{ns}) and (\ref{dnsk}) the scalar spectral index and its running could be achieved as
\begin{equation}
n_s(n, k, N)=\frac{{14 + k - 6n - 16N + 4nN}}{{( - 4 + n)(k + 4N)}}~.\label{nsN}
\end{equation}
and
\begin{equation}
\alpha_s(n, k, N)= - \frac{{4( - 14 + k( - 5 + n) + 6n)}}{{( - 4 + n){{(k + 4N)}^2}}}~.\label{dnsN}
\end{equation}
Also from Eq.(\ref{pt}), the tensor power spectrum in terms of the number of e-fold is obtained as
\begin{equation}
{{\cal P}_t}=\frac{{2{{(\frac{{{k^2}{M_p}^2}}{2} + 2k{M_p}^2N)}^{k/2}}{V_0}}}{{3{M_p}^4{\pi ^2}}}~. \label{ptN}
\end{equation}
Additionally, from Eq. (\ref{rratio}), the tensor to scalar ratio is obtained as
\begin{eqnarray}\nonumber
r{\rm{ }}(n, k, N) &=&\sqrt {\frac{{{{\rm{V}}_0}}}{3}}\\
&\times&  \frac{{{{(\frac{{{k^2}{M_p}^2+ 4k{M_p}^2N}}{2})}^{\frac{{3k + 6n - kn - 14}}{{2(8 - 2n)}}}}}}{{25 (8{k^2}{M_p})^{-1}{\pi ^2}\gamma^\ast }} a^{\frac{1}{n-4}} ~.\label{rNweak}
\end{eqnarray}
Now to make the constraints on free parameters of the model, we need to compare them  with observational data originated from Planck $2013$ and $2015$ data \cite{Ade:2013ktc,Ade:2013uln,Planck2015}. For this purpose,  we plot the free parameters $n-k$ diagram by virtue of Eqs.(\ref{rNweak}) and (\ref{nsN})  for the Confidence Levels (CLs)  68\% and 95\% CL allowed by Planck $2015$, TT, TE, EE+lowP data \cite{Planck2015}. This figure \ref{fignrweak} shows the regions in which the {pair} of free parameters, they can make a set of ${n, k}$, satisfy the constraints originated from Planck data.

\begin{figure}[ht]
\centering
\includegraphics[scale=.55]{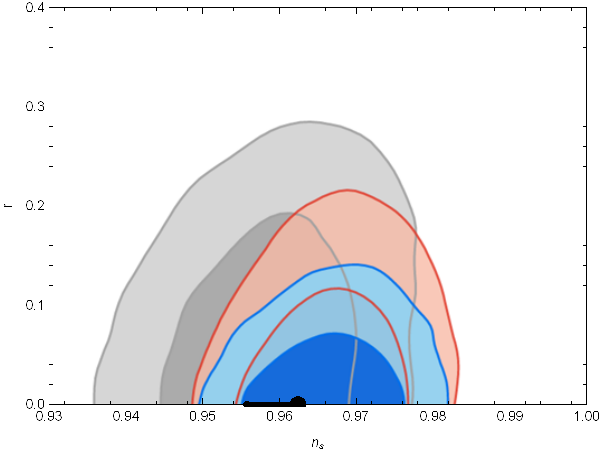}
\includegraphics[scale=.55]{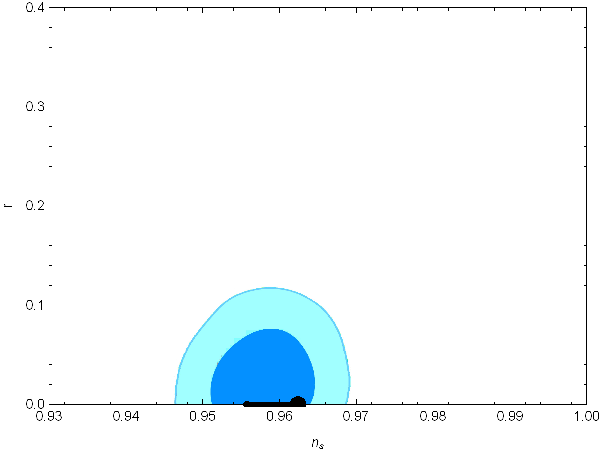}
\caption{{\it{
 The $r-n_s$ diagram shows  Prediction of
the model  in the weak regime for free parameters  $\alpha=70$, $k=6$,  $n=2$, $V_0=2.7\times10^{-25}$, $M_p=1$ and $N_e=60$,
in comparison to the observational data risen by Planck $2013$, $2015$ and $2018$. In the left panel, the likelihood  of  Planck 2013 are indicated with grey contours, Planck 2015 TT+lowP with red contours, and Planck 2015 TT,TE,EE+lowP with blue contours. And in the right panel, the results of Planck 2018 are indicated by dark and light blue colours referring $68\%$ and $95\%$ confidence levels respectively. In both figures the thick black lines
refers the predictions of theoretical results in which small and large circles are the values of $n_s$ at the number of e-folds $N=55,~N=65$ respectively.}}}
\label{fignsr}
\end{figure}

Now by means of an arbitrary set of the aforementioned pairs we can re-plot the  $r-n_s$ diagram as one of the most important {instrument} to test the accuracy of a theoretical model. So the figure \ref{fignsr} shows the diagram $r-n_s$ based on different types of data mentioned in the caption. Figure \ref{fignsr} indicates that our results are in a good agreement compared to the observations.

\begin{figure}[ht]
\centering
\includegraphics[scale=.53]{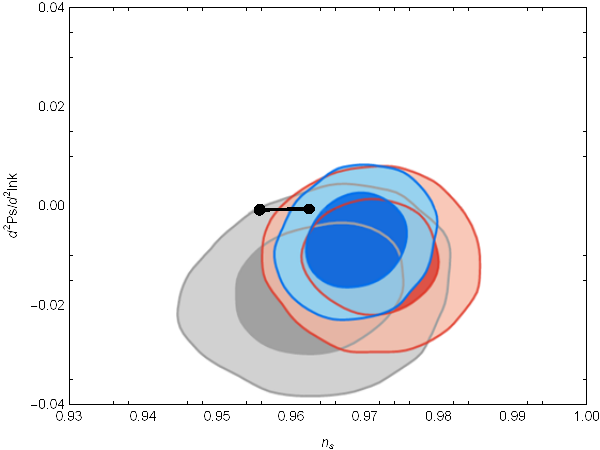}
\caption{{\it{
 In this figure the thick black line
depicts the predictions of our model in which small and large circles are the value of $n_s$ at the number of e-fold $N=55,~N=65$ respectively. To plot this shape we use the free parameters $k=6$,  $n=2$, $V_0=2.7\times 10^{-25}$, $M_p=1$, $N_e=65$ and $\alpha=70$.
In this figure  the grey contours indicates the likelihood  of  Planck 2013, Planck 2015 TT+lowP showed with red contours and the blue contours considered for Planck 2015 TT,TE,EE+lowP.}}}
\label{fignsdns1}
\end{figure}

The latest observational data suggest that the amplitude of
scalar perturbation at the horizon crossing is very close {to the amount of}
{${\cal P}_s=2.17\pm 0.1\times 10^{-9}$}, and the tensor-to-scalar
ratio  has an upper limit as $r < 0.11$ at 68\% { \cite{Planck2015}}.
In what following, by means of Eqs.(\ref{dnsN}) and (\ref{nsN}) {we  want to plot  the diagram of} running parameter, $d{n_s}/dN -
{n_s}$ and then compare the obtained results with the observations originated from Planck data.   Figure \ref{fignsdns1} which indicates  the prediction of this model  can lie {inside} the joint 68\% CL region of Planck 2015 TT, TE, EE+lowP data {\cite{Planck2015,Planck2018}}, and so could satisfy the  compatibility with observations.

\begin{figure}[ht]
  \centering
  \includegraphics[scale=.6]{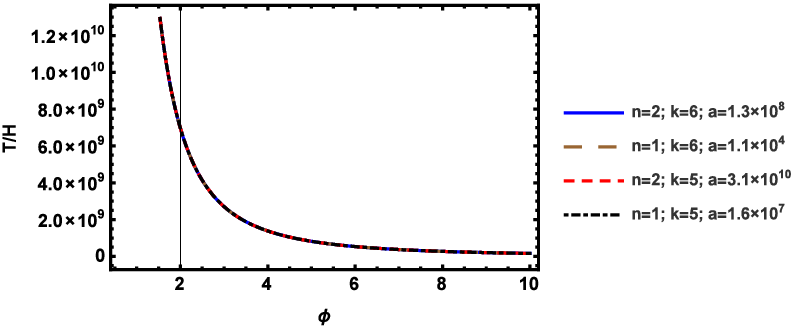}
  \caption{\emph{{The plot shows the ratio of the temperature to the Hubble parameter during the inflationary period of the  model in the weak dissipative regime versus the inflaton scalar field, $\phi$, for different values of $(n,k)$ and the parameter $a$. As one can see from the plots during inflation the temperature is larger than the Hubble parameter, and the condition $T>H$ is satisfied properly. To draw these figures we fixed the value of $V_0$ at $10^{-25}$}.}}\label{thweak}
\end{figure}

{An important feature of the warm inflationary model is that the thermal fluctuations overcome the quantum fluctuations, since the fluid temperature is bigger than the Hubble parameter , i.e. $T>H$. To receive a healthy warm inflation, this condition should be justified during the cosmological evolution. Fig. \ref{thweak} expresses the behavior of the ratio of the temperature to the Hubble parameter during such era. From the Figure \ref{thweak}, one can observe that the condition is satisfied during the inflationary phase of cosmological dynamics.}

\begin{figure}[ht]
  \centering
  \includegraphics[scale=.7]{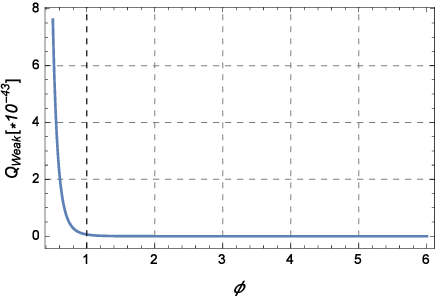}
  \caption{\emph{{{This figure indicates the ratio of the dissipation parameter,$\Gamma$, to the Hubble parameter during the inflationary period of the  model in the weak dissipative regime versus the inflaton scalar field, $\phi$, for the value of $(n=3,k=7)$ and the parameter $a=10^{-4}$. As observed from the plot during inflation the parameter $\Gamma$ is much less than the Hubble parameter, and the condition $Q=\Gamma/3H<<1$ is satisfied properly. To draw these figure we fixed the value of $V_0$ at $10^{-25}$}}.}}\label{Qweak}
\end{figure}

{Although it has postulated that $Q<<1$ in the weak regime, it should be controlled by virtue of the best fitted free parameters of the model. Hence one can observe from Fig. \ref{Qweak} this condition is satisfied properly.}

\section{Consistency Of Power-Law Potential In strong dissipative Regime }\label{strongpl}
\begin{figure}[ht]
\centering
\includegraphics[scale=.43]{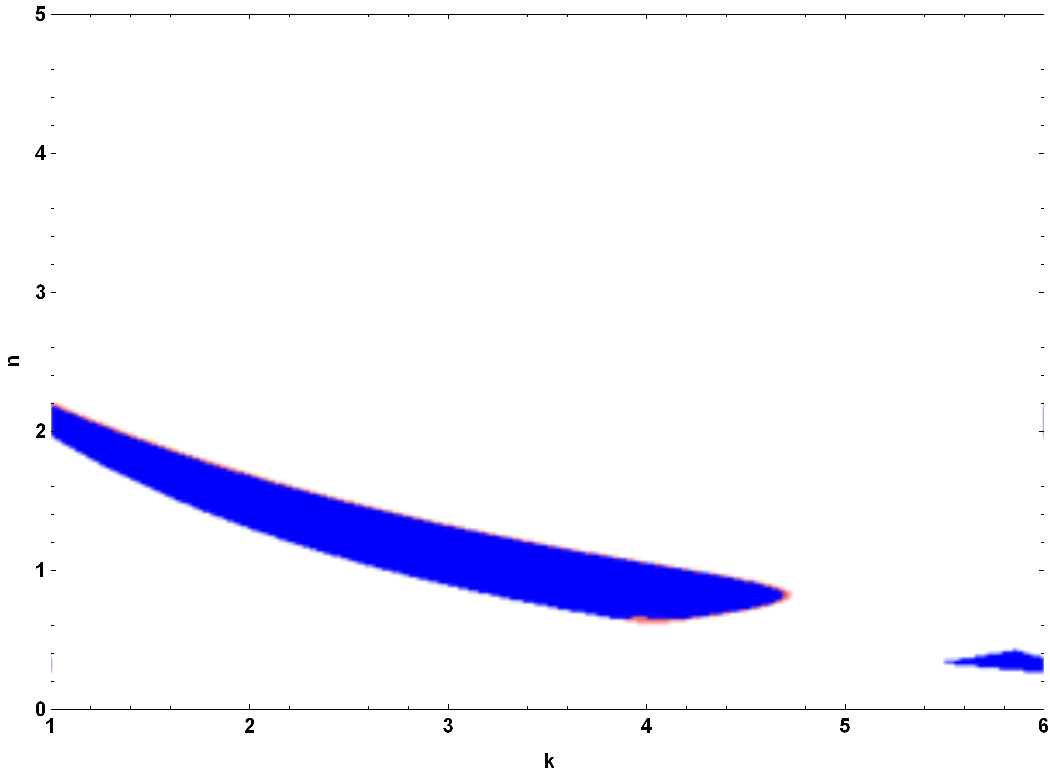}
\caption{{\it{
 This plot shows a set of $(n,k)$ pair that can give the best fitted values based on the  $(r-n_s)$ diagram originated from Planck $2015$ observational data. Here the dark blue color shows the results for $68\%$ CLs of Planck $2015$ data, and the light red color shows those, that put the result in $95\%$ CLs. To plot this diagram we considered $V_0=5.0\times10^{-20}$, and $M_p=1$.}}}
\label{StrongPlotNsraa}
\end{figure}

\begin{figure}[ht]
\centering
\includegraphics[scale=.61]{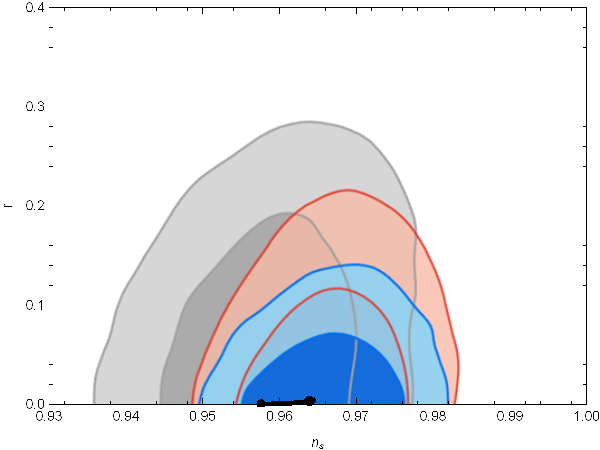}
\includegraphics[scale=.43]{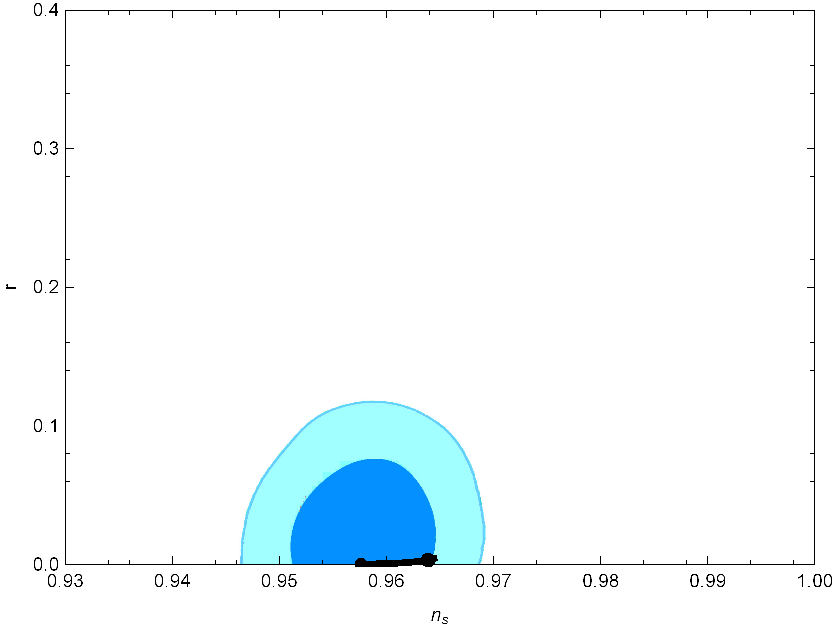}
\caption{{\it{
 The $r-n_s$ diagram shows  Prediction of
our theoretical results  in strong regime for free parameters  $\alpha=70$, $k=1$,  $n=4$ and $V_0=5\times10^{-20}$, $M_p=1$ and $N_e=65$,
in comparison to the observational data risen by Planck $2013$, $2015$ and 2018. In the left panel, the likelihood  of  Planck 2013 are indicated with grey contours, Planck TT+lowP with red contours, and Planck TT,TE,EE+lowP with blue contours. And in the right panel, the results of Planck 2018 are indicated by dark and light blue colours referring $68\%$ and $95\%$ confidence levels respectively. In both figures the thick black lines
refers the predictions of theoretical results in which small and large circles are the values of $n_s$ at the number of e-folds $N=55,~N=65$ respectively.
}}}
\label{fignsrstrong}
\end{figure}

\begin{figure}[ht]
\centering
\includegraphics[scale=.43]{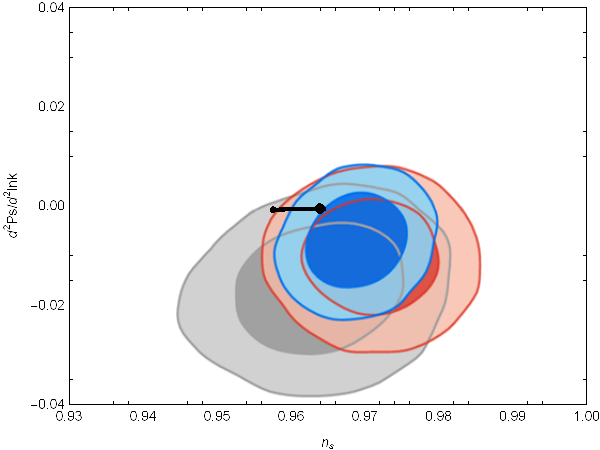}
\caption{{\it{
 This is the $d{n_s}/dN - {n_s}$ diagram which explains the running of the parameter ${n_s}$. In this figure the thick black line
depicts the predictions of our model in which small and large circles are the value of $n_s$ at the number of e-fold $N=55 and ~N=65$ respectively. To plot this result we use the free parameters $k=1$,  $n=4$,  and $V_0=5\times 10^{-20}$, $M_p=1$, $N_e=65$ and $\alpha=70$.
In this diagram  the grey contours indicated for likelihood  of  Planck 2013, Planck TT+lowP showed with red contours and the blue contours considered for Planck TT,TE,EE+lowP.y.}}}
\label{strongfignsdns}
\end{figure}

Here we want to continue our calculations {this time} for strong dissipation regimes namely $Q\gg1$. {Clearly} by considering Eqs. (\ref{Freq}), (\ref{phidot}), (\ref{Potential})
  and slow-roll approximation, the Friedman and evolution scalar field equations take the following form respectively,
  \begin{equation}
\ H^2=\frac{1}{3M^2_p}{\ V(\phi)}~, \label{WeakFreq}%
\end{equation}
and
\begin{equation}\label{phidotstrong}
 3H Q\dot \phi  + V_{,\phi} = 0~.
\end{equation}
Using the Eqs. (\ref{Gamma}), (\ref{rhor}),  (\ref{phidotstrong}) and by virtue of the above inflationary potential, i.e.  (\ref{Potential}), the temperature could be obtained as follows:
\begin{equation}\label{tempStrong}
T=\bar{\gamma} {\phi ^{\frac{{ - 3 + \frac{{3k}}{2} + n}}{{4 + n}}}},
\end{equation}
where \[\bar{\gamma}={(\frac{{{k^2}{M_p}\sqrt {V_0^3} }}{{2520\alpha }})^{\frac{1}{{4 + n}}}}{a^{\frac{{ - 1}}{{4 + n}}}} = \overline {{\gamma ^*}} {a^{\frac{{ - 1}}{{4 + n}}}}.\]
Now, by virtue  of the above equations the first slow-roll parameter  and dissipation term $Q$ are obtained respectively as bellow:
\begin{eqnarray}\label{epsilon-strong}
\epsilon= {(\frac{{k{M_P}}}{{\sqrt 2 \phi }})^2}~,
\end{eqnarray}
and
\begin{eqnarray}
Q = \frac{{{M_P}a{{\bar \gamma }^n}}}{{\sqrt {3{V_0}} }}{\phi ^{\frac{{4 - 6n - 2k + kn}}{{4 + n}}}}~.\label{srw}
\end{eqnarray}
For the strong regime usually inflation ends up when $\epsilon\approx Q$, {so }using Eq.(\ref{srw}) the scalar field related to such high friction eras is obtained as
\begin{equation}
{\phi _{end}} = {(\frac{{2a{{\bar \gamma }^n}}}{{ {M_p}{k^2}\sqrt {{3V_0}} }})^{ - \frac{{4 + n}}{{12 - 2k - 4n + kn}}}}~.\label{phiend}
\end{equation}
Additionally, from  Eq.(\ref{Nefold}) the value of the scalar field based on the horizon line crossing for perturbations is given by
\begin{equation}\label{PhiStrong}
\phi  = {(k{M_P}\sqrt {3{V_0}} {{\bar \gamma }^{( - n)}}{a^{ - 1}}[(2 + \lambda )N + \frac{k}{2}])^{\frac{1}{{2 + \lambda }}}}~,
\end{equation}
where $\lambda=\frac{{4 - 6n - 2k + kn}}{{4 + n}}.$ In this regime, similar to the  weak case,  to specify scalar power spectrum  we consider the Eq.(\ref{ps}) but here $\delta\phi^2\simeq k_F T/2\pi^2$  where $k_F=\sqrt{\Gamma H}$ and it can be appeared as follows {\cite{Herrera:2015aja,sayar}}:
\begin{equation}
{{\cal P}_s}=\frac{{25}}{{32\,{\pi ^2}}}{\left( {\frac{{4\,{\Gamma ^3}\,{V^{3/2}}}}{{{\alpha ^{1/3}}{3^{3/2}}{M_P}^3\,{{V'}^2}}}} \right)^{\frac{3}{4}}}~.\label{psVStrong}
\end{equation}
Now by substituting Eqs.(\ref{Gamma}, \ref{tempStrong}) and (\ref{PhiStrong}) into Eq.(\ref{psVStrong}) one obtains
\begin{equation}
\begin{array}{l}
{{\cal P}_s} =
\frac{{25{{\left( {\frac{{({{\overline {{\gamma ^*}}  }})^{3n}}}{{{k^2}M_P^3\sqrt {{V_0}} {\alpha ^{1/3}}}}} \right)}^{\frac{3}{4}}}}}{{{3^{1/8}}24{\pi ^2}\sqrt 2 }} \times \\
{\left( {{{\left[ {\sqrt 3 k{M_P}\sqrt {{V_0}}~ {{\overline {{\gamma ^*}}  }}^{( - n)}\left( {N\left( {2 + \lambda } \right) + \frac{k}{2}} \right)} \right]}^{\frac{1}{{2 + \lambda }}}}} \right)^{\frac{{3\beta }}{4}}}{a^{{3\psi}} }
 \end{array}\label{psNStrong}
\end{equation}
where $\beta  = \frac{{ - 2k + 8 + 4kn - 3n}}{{4 + n}}$ and $$\psi=\frac{{ -4k - kn  + 15n + 24}}{{( - 4 + k){n^2} + 2kn - 8k + 4n + 48}}~.$$
 To make a better agreement between observational data and theoretical results of our investigation we can follow the approach which { done} in weak regime. Thence, if we want to precisely consider the Planck {results} for ${{\cal P}_s}$, the parameter $a$ based on ${{\cal P}_s}$ and the free parameters of the model can be achieved as
 \begin{equation}
\begin{array}{l}\label{a-Strong}
 a(n,k,N) = {\left( {\frac{{{3^{1/8}}24{\pi ^2}\sqrt 2 }}{{25{{\left( {\frac{{{{(\overline {{\gamma ^*}} )}^{3n}}}}{{{k^2}M_P^3\sqrt {{V_0}} {\alpha ^{1/3}}}}} \right)}^{\frac{3}{4}}}}}} \right)^{\frac{1}{{3\psi }}}} \times\\
  {\left( {{{\left[ {\sqrt 3 k{M_P}\sqrt {{V_0}} {{\overline {{\gamma ^*}} }^{( - n)}}\left( {N\left( {2 + \lambda } \right) + \frac{k}{2}} \right)} \right]}^{\frac{{ - 3\beta }}{{8 + 4\lambda }}}}} \right)^{\frac{1}{{3\psi }}}}{{{\cal P}_s}}^{\frac{1}{{3\psi }}}~.
  \end{array}
\end{equation}
Besides, from Eqs.(\ref{ns}) and (\ref{dnsk}) the scalar spectral index and its running can be obtained as follow
\begin{eqnarray}\nonumber
n_s(n, k, N)=\\
&1 + \frac{{ - 24 + k\left( {6 - 12n} \right) + 9n}}{{2\left( { - 8\left( { - 3 + n} \right)N + k\left( {4 + n - 4N + 2nN} \right)} \right)}},\label{nsNStrong}
\end{eqnarray}
and
\begin{eqnarray}\label{dnsNStrong}\nonumber
\alpha _s(n, k, N)=\\ \nonumber
&-\frac{{3\left( { - 4\left( { - 3 + n} \right) + k\left( { - 2 + n} \right)} \right)\left( {8 - 3n + k\left( { - 2 + 4n} \right)} \right)}}{{{{\left( { - 8\left( { - 3 + n} \right)N + k\left( {4 + n - 4N + 2nN} \right)} \right)}^2}}}~.\\
\end{eqnarray}
From Eqs.(\ref{pt}) and (\ref{PhiStrong}), the tensor power spectrum in terms of the number of e-fold gives
\begin{eqnarray}\label{ptNStrong}
\begin{array}{l}
{{\cal P}_t}(n, k, N) =\frac{{2{V_0}}}{{3{\pi ^2}M_P^4}} \times \\
\,{\left( {\frac{{ {\sqrt 3 k{M_P}\sqrt {{V_0}}~ {{\overline {{\gamma ^*}} }^{( - n)}}\left( {N\left( {2 + \lambda } \right) + \frac{k}{2}} \right)}}}{{{a^{\frac{4}{{n + 4}}}}}}} \right)^{\frac{k}{{2 + \lambda }}}}~,
\end{array}
\end{eqnarray}
in addition, from Eqs. (\ref{rratio}, \ref{ptNStrong}) and (\ref{psNStrong}), the tensor to scalar ratio for strong regime is expressed as follows:
\begin{equation}
\begin{array}{l}
r(n, k, N) = \frac{{16\sqrt 2  \times {3^{1/8}}{V_0}}}{{25M_P^4{{\left( {\frac{{{{\overline {{\gamma ^*}} }^{(3n)}}}}{{{k^2}M_P^3\sqrt {{V_0}} {\alpha ^{1/3}}}}} \right)}^{3/4}}}}{a^{(\frac{{ - 4k}}{{(n + 4)(2 + \lambda )}} - 3\psi )}} \times \\
{\left( {\sqrt 3 k{M_P}\sqrt {{V_0}} ~{{\overline {{\gamma ^*}} }^{( - n)}}\left( {N\left( {2 + \lambda } \right) + \frac{k}{2}} \right)} \right)^{\frac{{4k - 3\beta }}{{8 + 4\lambda }}}}~.
 \end{array}\label{rNStrong}
\end{equation}
Now we  should evaluate consistency of our model with observations  originated from Planck $2013$ and $2015$ data \cite{Ade:2013ktc,Ade:2013uln,Planck2015}. Similar to the weak case, we can plot the regions in which a pair of our free parameters $(n, k) $  could satisfy the best concord between theory and observations for $r-n_s$ parameters. Such results can be found in figure \ref{StrongPlotNsraa}, in which to fix our results we use the values $V_0=5.0\times10^{-20}$, and $M_p=1$. {After that, } by means of Eqs.(\ref{nsNStrong}) and (\ref{rNStrong}) we can consider an arbitrary pair of free parameters as $(n=4, k=1)$ to investigate the results based on usual $r-n_s$ diagram. The investigation resulted in figure \ref{fignsrstrong} and one can observe that our best fitted parameters are in a good agreement compared to Planck results  \cite{Planck2015}. Additionally, to investigate the behaviour of the running parameter we can use Eq.(\ref{dnsNStrong}), and likelihood diagram from Planck results to compare our model, see figure \ref{strongfignsdns}.

Interestingly, our investigation confirms that although one of the best power law potentials to run the inflation is the quartic ones, another { powers of power law potential} could be considered as long as we choose the best fitted parameters in a {correct} way. So we can confirm that for both regimes the power law potentials in the framework of warm inflation are in good agreement compared to the planck results.

\begin{figure}[ht]
  \centering
  \includegraphics[scale=.6]{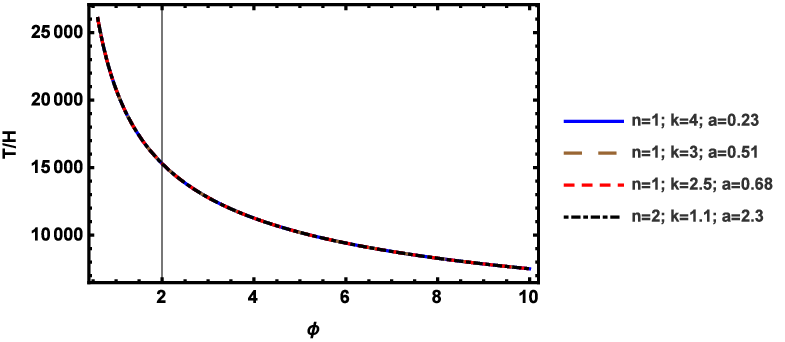}
  \caption{\emph{{The ratio of the temperature to the Hubble parameter during the inflationary period of the  model in the weak dissipative regime versus the inflaton scalar field, $\phi$, for different values of $(n,k)$ and the parameter $a$. As one can see from the plots during inflation the temperature is larger than the Hubble parameter, and the condition $T>H$ is satisfied properly. To draw these figures we fixed the value of $V_0$ at $10^{-20}$.}}}\label{thstrong}
\end{figure}

{As explained for the weak regime, an important property of the warm inflationary paradigm is that the thermal fluctuations overcome the quantum fluctuations, since the fluid temperature is bigger than the Hubble parameter. To derive a healthy warm inflation, this condition should be justified during the primordial evolutions. Fig. \ref{thstrong} expresses the behavior of the ratio of the temperature to the Hubble parameter during such era. From the Figure \ref{thstrong}, one can observe that the condition is satisfied during the inflationary phase of cosmological dynamics.}

\begin{figure}[ht]
  \centering
  \includegraphics[scale=.7]{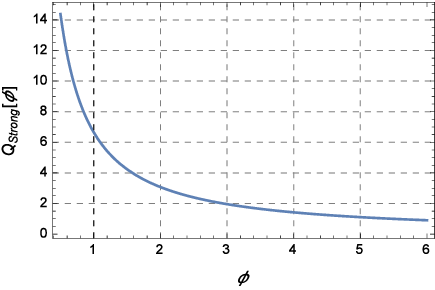}
  \caption{\emph{{{This figure indicates the ratio of the dissipation parameter,$\Gamma$, to the Hubble parameter during the inflationary period of the  model in the strong dissipative regime versus the inflaton scalar field, $\phi$, for the value of $(n=1.3,k=3)$ and the parameter $a=0.2$. As observed from the plot during inflation the parameter $\Gamma$ is  large than the Hubble parameter, and the condition $Q=\Gamma/3H>1$ is satisfied properly. To draw these figure we fixed the value of $V_0$ at $10^{-20}$}}.}}\label{QStrong}
\end{figure}

{In the strong regime  it has assumed that $Q>1$ , and it must control by means of the best fitted free parameters of the model. Therefore, one can see from the plot  Fig. \ref{QStrong} that this condition is satisfied properly.}

\section{Conclusions}\label{seccon}
By introducing a general form for the power-law potentials and the dissipation coefficient {in } the frame work of warm inflationary scenario the constraints originated from Planck $2013$ and $2015$ have been studied. This study was separated into two well-know regimes, based on the friction of the primordial soup, namely weak and strong regimes. Besides, for a quasi-static universe the slow-roll constraints have been considered. After these simplification approaches,  at first the different inflationary parameters such as  power spectrum of scalar and tensor perturbations, scalar and tensor spectral indices, running of them and tensor-to-scalar ratio, in both weak and strong regimes {were} been calculated.
To visualize how the aforementioned parameters  can satisfy a concord between theoretical and observations, we {have considered }a way in which a region of free parameters could do it, instead of just one set of the free parameters. In other word, by virtue of $r-n_s$ diagram, for both regimes, and fixing one of the free parameters the likelihood of free parameters  $(n, k)$ {were} been plotted. The results have been appeared in figures \ref{fignrweak}   and \ref{StrongPlotNsraa}   for weak and strong regimes respectively. In these figures two important CLs namely $68\%$ and $95\%$ for more clearance have been distorted  by means of  blue and red in colors. So we found the best fitted regions of free parameters that could make a good estimations for the theory. To examine our estimations an arbitrary pair of free parameters for both regimes have been illustrated. At first, and to avoid any confusions let us turn on to the weak regime investigations. For this regime the pair $(n=2, k=6)$ with $V_0=2.7\times 10^{-25}$ have been selected. Based on this selection, the diagram {$r-n_s$} and its running based on Planck data have been plotted, see figures \ref{fignsr} and \ref{fignsdns1}. These figures {have } shown that our approach and also our results are in a good consistency in compared with the observations originated from Planck $2013$ and $2015$ data sets. Although, for weak regime we considered the power {of variable}  of the potential as $k=6$, as well it could be observed for the quadratic or quartic powers there are no any failures. For the coefficient $V_0$ and other constants of the model our results can be justified compared to literature. {We should add here, we can consider different values for the potential coefficient, i.e. $V_0$ in which  the regions at  figures \ref{fignrweak}   and \ref{StrongPlotNsraa}  showon a little bit changes, so we avoided  to  the re-plotting  them}. Now we can turn our attention to the strong regime. In this regime because of the presence of the parameter $Q$, the situation is completely different and also maybe a little bit difficult comparing with weak case. For strong dissipation regime, the arbitrary pair of free parameters has been considered as $(n=1, k=4)$ with $V_0=5.0\times 10^{-20}$. For this regime our freedom in selection of free parameter is more less than weak case, because of the smaller region for best fitted free parameters. Again, the accuracy of the obtained pairs for free parameters can be examined by means of the criterions {arisen} from Planck results. We did it by plotting the { $r-n_s$} and its running and fortunately our results  have been in a good agreement compared to the observations, Planck data. Our claims can be justified by looking at figures \ref{fignsrstrong}   and \ref{strongfignsdns} for more investigations. We can continue this study  about the relation between scalar field obtained here and also the scalar field obtained conformally to see the effects of amplitude of gravitational  waves on the behaviour of inflationary observables, might in a new project.
{It has showed that Figs. \ref{thweak} and  \ref{thstrong} can  justify the satisfactory behavior of the ratio of the temperature to the Hubble parameter during inflationary epoch for both weak and strong cases, i.e. $T>H$.}
{To derive a healthy and self-consistent  warm inflationary model, besides the whole above conditions one hast to control the constraints on the dissipation function for the both of the weak and the strong regimes. In doing so, we have plotted two figures which showed a satisfactory concord between theoretical results of this work and observational Planck data. The Figs. \ref{Qweak} and \ref{QStrong} have indicated the ratio of the dissipation parameter,$\Gamma$, to the Hubble parameter during the inflationary period of the  model  and were satisfied properly.}

\section{Acknowledgement}
The authors would like to thank A. Mohammadi for the serious discussions on the likelihood plots resulted in regional best fitted parameters. HS thanks A. Starobinsky for very constructive discussions about inflation during Helmholtz International Summer School  2019 in Russia. He grateful G. Ellis, A. Weltman, and UCT for arranging his short visit, and for enlightening discussions about cosmological fluctuations and perturbations for both large and local scales. He also thanks  H. Firouzjahi for constructive discussions about inflation and perturbations. His special thanks go to his wife E. Avirdi for her patience during our stay in South Africa.

\end{document}